\documentclass[referee]{aa}
\usepackage{graphicx}
\usepackage{natbib}

\begin{document}

\title{Galactic orbits of stars with planets \thanks{Tables 1, 2, and 4 
are only available in electronic form at the CDS via
anonymous FTP to cdsarc.u-strasbg.fr (130.79.128.5)}}

\author{
M. Barbieri\inst{1}
\and 
R.G. Gratton\inst{2} 
}

\authorrunning{M. Barbieri \& R.G. Gratton}
\titlerunning{Galactic orbits of stars with planets}

\offprints{M. Barbieri, \email{barmau@pd.astro.it}}

\institute{ 
Dipartimento di Astronomia, Universit\`a di Padova, Vicolo dell'Osservatorio, 2,
I-35122 Padova, ITALY, \email{barmau@pd.astro.it}\\
\and
Osservatorio Astronomico di Padova, Vicolo dell'Osservatorio, 5,
I-35122 Padova, ITALY, \email{gratton@pd.astro.it}\\
}

\date{Received 7 August 2001/Accepted 5 December 2001}

\abstract{
We have reconstructed the galactic orbits of the parent stars of exoplanets.
For comparison, we have recalculated the galactic orbits of stars from the
\cite{edva93} catalog. A comparison between the two samples
indicates that stars with planets are not kinematically peculiar. At each
perigalactic distance 
stars with planets have a metallicity systematically larger
than the average for the comparison sample. We argue that this result favors
scenarios where the presence of planets is the cause of the higher metallicity
of stars with planets.
}

\maketitle

\section{Introduction}

Spectroscopic analysis of parent stars of exoplanets had shown that these stars
are more metal-rich than field stars (\citealt{gon96,but00}). Two
scenarios had been proposed to explain the high metallicity: (i) During the
build-up of planets, the gravitational interaction among them (or with the
disk) injects some objects in high-eccentricity orbits that can intersect the
surface of the host star. If a sufficient number of these objects are captured
by the star, then the photospheric metallicity will be enhanced by the
dissolution of the planet. (ii) Planet formation is enhanced by the high
metallicity of the parent protostar nebula.

If there is a correlation between metal abundances and other properties
of the stars that should have no impact on the formation of planets
(e.g. their kinematics), we may expect systematic differences between stars 
with planets and without planets, in the first scenario, while there
should be no difference in the second scenario.

In this paper we study the second hypothesis. Following the work by 
\cite{asa91}, using stellar parallaxes and proper motions from Hipparcos,
we have reconstructed the galactic orbit of the parent stars of exoplanets.
For comparison, we have recalculated the galactic orbit from the \cite{edva93} 
catalogue, adopting a solar galactocentric distance of 8.5 kpc instead
of the 8.0 kpc used by Edvardsson. A comparison between the two samples indicates 
that they are quite similar, and therefore that the stars with planets are not
kinematically peculiar.

\section{The samples}

The stars with planets (SWP) are stars similar to the Sun; this peculiar
characteristic is an intrinsic selection effect of discovery methods of radial
velocities. The SWP sample we consider includes 58 stars, located within 70 pc
of the Sun. Most of them are single; only a few stars are wide
binaries, with a separation larger than 700 AU between the two components.
We have retrieved the list of SWP from the Extrasolar Planets Encyclopaedia at 
{\tt http://www.obspm.fr/encycl/encycl.html} maintained by J. Schneider.
The samples considered by the Geneva and Marcy and Butler groups 
(that have discovered the vast majority of extrasolar planets discovered 
so far, and carry most of the weight in our discussion) 
do not contain any kinematic bias.
The samples considered by these two groups preferentially include 
low activity stars, excluding in this manner most of the stars with 
ages $<$ 2 Gyr.

The comparison sample is the same sample used by \cite{edva93} to study the
chemical evolution of the galactic disk. The sample of Edvardsson is composed
of 189 nearby ($d<$ 80 pc) F and G disk stars. The stars are selected from the
\cite{ols88} catalogue, in the metal abundance range $-1.1 < {\rm [Me/H]} <
0.3$; they are brighter than $V \simeq 8.3$. 
Only stars that have evolved off the Zero Age Main Sequence by more than 0.4 magnitude
were considered. This ensures that their sample does not contain stars
younger than 1.5 Gyr, and only very few younger than 2 Gyr (see their table 11).
In this respect, the selection criteria are very similar to those considered
for SWP. Due to the selection criteria
used by Edvardsson et al. in their sample is not representative of the distribution of
stars in the solar neighborhood with metallicity because it over-represents
low-metallicity stars with respect to solar-metallicity stars. However, the
sample may be used to study the dynamical properties of stars in the solar
neighborhood in a given abundance range because no kinematical selection
criterion was adopted.

Nine stars are in common between the two samples: HD 6434, HD 9826, HD 19994,
HD 89744, HD 95128, HD 114762, HD 143761, HD 169830, HD 217014.

\subsection{Kinematical data}

All the SWP (except one: BD-10$^\circ$ 3166) have been observed by {\it Hipparcos}
satellites. Using SIMBAD we have retrieved parallaxes and proper motion data
from the {\it Hipparcos} catalog. The radial velocity are taken from the
initial reports if available, we have taken the values from otherwise \cite{bfi00},
\cite{egg98a} and \cite{car94} catalogs. The majority of these stars are
classified as ``High Proper Motion stars''. Kinematical data for the SWP are
given in Table~\ref{t:tab1} (only available in electronic form).

\begin{table*}
\centering
\caption{Kinematical data for stars with planets (in electronic form)}
\scriptsize
\begin{tabular}{lrrrrrrrrrrrrrrr}
\hline
Star            &      $p$        &    $V_r$     & $\mu_\alpha$   &  $\mu_\delta$  &  $\alpha$        & $\delta$  \\
                &      mas        &    Km/s      &   mas/yr       &    mas/yr      &   2000.0         &  2000.0   \\
\hline
HD 1237    &$   56.76  $&$  -5.808 $&$  433.88 $&$  -57.91 $& 00$^h$ 16$^m$ 12.68$^s$ & -79$^\circ$ 51$^{'}$ 04.3$^{"}$\\
HD 6434    &$   24.80  $&$  22.962 $&$ -168.97 $&$ -527.70 $& 01$^h$ 04$^m$ 40.15$^s$ & -39$^\circ$ 29$^{'}$ 17.6$^{"}$\\
HD 8574    &$   22.65  $&$  18.864 $&$  252.59 $&$ -158.59 $& 01$^h$ 25$^m$ 15.52$^s$ &  28$^\circ$ 34$^{'}$ 00.1$^{"}$\\
HD 9826    &$   74.25  $&$   -27.7 $&$ -172.57 $&$ -381.03 $& 01$^h$ 36$^m$ 47.84$^s$ &  41$^\circ$ 24$^{'}$ 19.7$^{"}$\\
HD 10697   &$   30.71  $&$   -44.8 $&$  -45.05 $&$ -105.39 $& 01$^h$ 44$^m$ 55.82$^s$ &  20$^\circ$ 04$^{'}$ 59.3$^{"}$\\
HD 12661   &$   26.91  $&$   -52.2 $&$ -107.81 $&$ -175.26 $& 02$^h$ 04$^m$ 34.29$^s$ &  25$^\circ$ 24$^{'}$ 51.5$^{"}$\\
HD 13445   &$   91.63  $&$   56.57 $&$ 2092.59 $&$  654.49 $& 02$^h$ 10$^m$ 25.93$^s$ & -50$^\circ$ 49$^{'}$ 25.4$^{"}$\\
HD 16141   &$   27.85  $&$   -51.5 $&$ -156.89 $&$ -437.07 $& 02$^h$ 35$^m$ 19.93$^s$ & -03$^\circ$ 33$^{'}$ 38.2$^{"}$\\
HD 17051   &$   58.00  $&$    15.5 $&$  333.72 $&$  219.21 $& 02$^h$ 42$^m$ 33.47$^s$ & -50$^\circ$ 48$^{'}$ 01.1$^{"}$\\
HD 19994   &$   44.69  $&$  19.331 $&$  193.43 $&$  -69.23 $& 03$^h$ 12$^m$ 46.44$^s$ & -01$^\circ$ 11$^{'}$ 46.0$^{"}$\\
HD 22049   &$  310.74  $&$    16.3 $&$ -976.36 $&$   17.98 $& 03$^h$ 32$^m$ 55.84$^s$ & -09$^\circ$ 27$^{'}$ 29.7$^{"}$\\
HD 27442   &$   54.84  $&$    29.3 $&$  -47.99 $&$ -167.81 $& 04$^h$ 16$^m$ 29.03$^s$ & -59$^\circ$ 18$^{'}$ 07.8$^{"}$\\
HD 28185   &$   25.28  $&$  50.246 $&$   80.85 $&$  -60.29 $& 04$^h$ 26$^m$ 26.32$^s$ & -10$^\circ$ 33$^{'}$ 02.9$^{"}$\\
HD 37124   &$   30.08  $&$   -19.0 $&$  -79.75 $&$ -419.96 $& 05$^h$ 37$^m$ 02.49$^s$ &  20$^\circ$ 43$^{'}$ 50.8$^{"}$\\
HD 38529   &$   23.57  $&$    30.0 $&$  -80.05 $&$ -141.79 $& 05$^h$ 46$^m$ 34.91$^s$ &  01$^\circ$ 10$^{'}$ 05.5$^{"}$\\
HD 46375   &$   29.93  $&$     4.0 $&$  114.24 $&$  -96.79 $& 06$^h$ 33$^m$ 12.62$^s$ &  05$^\circ$ 27$^{'}$ 46.5$^{"}$\\
HD 50554   &$   32.23  $&$  -3.861 $&$  -37.29 $&$  -96.36 $& 06$^h$ 54$^m$ 42.83$^s$ &  24$^\circ$ 14$^{'}$ 44.0$^{"}$\\
HD 52265   &$   35.63  $&$    53.6 $&$ -115.76 $&$   80.35 $& 07$^h$ 00$^m$ 18.04$^s$ & -05$^\circ$ 22$^{'}$ 01.8$^{"}$\\
HD 74156   &$   15.49  $&$   3.813 $&$   24.96 $&$ -200.48 $& 08$^h$ 42$^m$ 55.12$^s$ &  04$^\circ$ 34$^{'}$ 41.2$^{"}$\\
HD 75289   &$   34.55  $&$   9.258 $&$  -20.50 $&$ -227.68 $& 08$^h$ 47$^m$ 40.39$^s$ & -41$^\circ$ 44$^{'}$ 12.4$^{"}$\\
HD 75732   &$   79.80  $&$    27.8 $&$ -485.46 $&$ -234.40 $& 08$^h$ 52$^m$ 35.81$^s$ &  28$^\circ$ 19$^{'}$ 50.9$^{"}$\\
HD 80606   &$   17.13  $&$   3.768 $&$   46.98 $&$    6.92 $& 09$^h$ 22$^m$ 37.57$^s$ &  50$^\circ$ 36$^{'}$ 13.4$^{"}$\\
HD 82943   &$   36.42  $&$   8.060 $&$    2.38 $&$ -174.05 $& 09$^h$ 34$^m$ 50.74$^s$ & -12$^\circ$ 07$^{'}$ 46.4$^{"}$\\
HD 83443   &$   22.97  $&$  28.917 $&$   22.35 $&$ -120.76 $& 09$^h$ 37$^m$ 11.83$^s$ & -43$^\circ$ 16$^{'}$ 19.9$^{"}$\\
HD 89744   &$   25.65  $&$     6.5 $&$ -120.17 $&$ -138.60 $& 10$^h$ 22$^m$ 10.56$^s$ &  41$^\circ$ 13$^{'}$ 46.3$^{"}$\\
HD 92788   &$   30.94  $&$    -4.0 $&$  -12.63 $&$ -222.75 $& 10$^h$ 42$^m$ 48.53$^s$ & -02$^\circ$ 11$^{'}$ 01.5$^{"}$\\
HD 95128   &$   71.04  $&$    12.0 $&$ -315.92 $&$   55.15 $& 10$^h$ 59$^m$ 27.97$^s$ &  40$^\circ$ 25$^{'}$ 48.9$^{"}$\\
HD 106252  &$   26.71  $&$  15.481 $&$   23.77 $&$ -279.41 $& 12$^h$ 13$^m$ 29.51$^s$ &  10$^\circ$ 02$^{'}$ 29.9$^{"}$\\
HD 108147  &$   25.93  $&$  -5.065 $&$ -181.60 $&$  -60.80 $& 12$^h$ 25$^m$ 46.27$^s$ & -64$^\circ$ 01$^{'}$ 19.5$^{"}$\\
HD 114762  &$   24.65  $&$    49.3 $&$ -582.68 $&$   -1.98 $& 13$^h$ 12$^m$ 19.74$^s$ &  17$^\circ$ 31$^{'}$ 01.6$^{"}$\\
HD 117176  &$   55.22  $&$     5.2 $&$ -234.81 $&$ -576.19 $& 13$^h$ 28$^m$ 25.81$^s$ &  13$^\circ$ 46$^{'}$ 43.6$^{"}$\\
HD 120136  &$   64.12  $&$   -15.8 $&$ -480.34 $&$   54.18 $& 13$^h$ 47$^m$ 15.74$^s$ &  17$^\circ$ 27$^{'}$ 24.9$^{"}$\\
HD 121504  &$   22.54  $&$  19.548 $&$ -250.55 $&$  -84.02 $& 13$^h$ 57$^m$ 17.24$^s$ & -56$^\circ$ 02$^{'}$ 24.2$^{"}$\\
HD 130322  &$   33.60  $&$ -12.504 $&$ -129.60 $&$ -140.79 $& 14$^h$ 47$^m$ 32.73$^s$ & -00$^\circ$ 16$^{'}$ 53.3$^{"}$\\
HD 134987  &$   38.98  $&$     5.2 $&$ -399.01 $&$  -75.10 $& 15$^h$ 13$^m$ 28.67$^s$ & -25$^\circ$ 18$^{'}$ 33.6$^{"}$\\
HD 141937  &$   29.89  $&$  -2.994 $&$   97.12 $&$   24.00 $& 15$^h$ 52$^m$ 17.55$^s$ & -18$^\circ$ 26$^{'}$ 09.8$^{"}$\\
HD 143761  &$   57.38  $&$    18.0 $&$ -196.88 $&$ -773.00 $& 16$^h$ 01$^m$ 02.66$^s$ &  33$^\circ$ 18$^{'}$ 12.6$^{"}$\\
HD 145675  &$   55.11  $&$ -13.842 $&$  132.52 $&$ -298.38 $& 16$^h$ 10$^m$ 24.31$^s$ &  43$^\circ$ 49$^{'}$ 03.5$^{"}$\\
HD 160691  &$   65.46  $&$    -9.0 $&$  -15.06 $&$ -191.17 $& 17$^h$ 44$^m$ 08.70$^s$ & -51$^\circ$ 50$^{'}$ 02.6$^{"}$\\
HD 168443  &$   26.40  $&$   -49.0 $&$  -92.15 $&$ -224.16 $& 18$^h$ 20$^m$ 03.93$^s$ & -09$^\circ$ 35$^{'}$ 44.6$^{"}$\\
HD 168746  &$   23.19  $&$ -25.645 $&$  -22.13 $&$  -69.23 $& 18$^h$ 21$^m$ 49.78$^s$ & -11$^\circ$ 55$^{'}$ 21.7$^{"}$\\
HD 169830  &$   27.53  $&$ -17.215 $&$   -0.84 $&$  -15.16 $& 18$^h$ 27$^m$ 49.48$^s$ & -29$^\circ$ 49$^{'}$ 00.7$^{"}$\\
HD 177830  &$   16.94  $&$   -72.3 $&$  -40.68 $&$ -051.84 $& 19$^h$ 05$^m$ 20.77$^s$ &  25$^\circ$ 55$^{'}$ 14.4$^{"}$\\
HD 178911  &$   20.42  $&$ -40.432 $&$   47.12 $&$  194.51 $& 19$^h$ 09$^m$ 04.38$^s$ &  34$^\circ$ 36$^{'}$ 01.6$^{"}$\\
HD 179949  &$   36.97  $&$   -25.5 $&$  114.78 $&$ -101.81 $& 19$^h$ 15$^m$ 33.23$^s$ & -24$^\circ$ 10$^{'}$ 45.7$^{"}$\\
HD 186427  &$   46.70  $&$   -27.5 $&$ -135.15 $&$ -163.53 $& 19$^h$ 41$^m$ 51.97$^s$ &  50$^\circ$ 31$^{'}$ 03.1$^{"}$\\
HD 187123  &$   20.87  $&$   -17.5 $&$  143.13 $&$ -123.23 $& 19$^h$ 46$^m$ 58.11$^s$ &  34$^\circ$ 25$^{'}$ 10.3$^{"}$\\
HD 190228  &$   16.10  $&$ -50.218 $&$  104.91 $&$  -69.85 $& 20$^h$ 03$^m$ 00.77$^s$ &  28$^\circ$ 18$^{'}$ 24.7$^{"}$\\
HD 192263  &$   50.27  $&$ -10.817 $&$  -63.37 $&$  262.26 $& 20$^h$ 13$^m$ 59.85$^s$ & -00$^\circ$ 52$^{'}$ 00.8$^{"}$\\
HD 195019  &$   26.77  $&$   -93.1 $&$  349.49 $&$  -56.85 $& 20$^h$ 28$^m$ 18.64$^s$ &  18$^\circ$ 46$^{'}$ 10.2$^{"}$\\
HD 209458  &$   21.24  $&$-14.7652 $&$   28.90 $&$  -18.37 $& 22$^h$ 03$^m$ 10.77$^s$ &  18$^\circ$ 53$^{'}$ 03.5$^{"}$\\
HD 210277  &$   46.97  $&$   -21.1 $&$  085.48 $&$ -449.83 $& 22$^h$ 09$^m$ 29.87$^s$ & -07$^\circ$ 32$^{'}$ 55.2$^{"}$\\
HD 213240  &$   24.54  $&$  -0.458 $&$ -135.16 $&$ -194.06 $& 22$^h$ 31$^m$ 00.37$^s$ & -49$^\circ$ 25$^{'}$ 59.8$^{"}$\\
HD 217014  &$   65.10  $&$   -33.6 $&$  208.07 $&$   60.96 $& 22$^h$ 57$^m$ 27.98$^s$ &  20$^\circ$ 46$^{'}$ 07.8$^{"}$\\
HD 217107  &$   50.71  $&$   -14.0 $&$   -6.05 $&$  -16.03 $& 22$^h$ 58$^m$ 15.54$^s$ & -02$^\circ$ 23$^{'}$ 43.4$^{"}$\\
HD 222582  &$   23.84  $&$         $&$ -145.41 $&$ -111.10 $& 23$^h$ 41$^m$ 51.53$^s$ & -05$^\circ$ 59$^{'}$ 08.7$^{"}$\\
BD-10 3166 &$ (\sim 10)$&$    26.4 $&$ -183.00 $&$   -4.80 $& 10$^h$ 58$^m$ 28.78$^s$ &  10$^\circ$ 46$^{'}$ 13.4$^{"}$\\
GJ 876     &$  212.69  $&$  -1.902 $&$  960.31 $&$ -675.61 $& 22$^h$ 53$^m$ 16.73$^s$ & -14$^\circ$ 15$^{'}$ 49.3$^{"}$\\
\hline
\normalsize
\end{tabular}
\label{t:tab1}
\end{table*}

The Edvardsson catalog does not contain data about the proper motion of the
stars: proper motions and the radial velocities for this second sample
were retrieved using SIMBAD.

\subsection{Abundances of the stars}

Some papers have been dedicated to the spectroscopic analysis of SWP: 
\cite{gon98}, \cite{gva98}, \cite{gon99}, \cite{gws99}, 
\cite{gla00}, \cite{gon01}, \cite{sim00}, \cite{nae01}.
From these works we notice that on average SWP are more metal-rich than
field stars. We have taken the metallicity [Fe/H] from these papers. Data
are unavailable for a few stars; in these cases they were taken from
\cite{v58} and \cite{cay01}.

The Edvardsson catalog contains [Fe/H] values for the stars.

While there may be systematic offsets between these different determinations,
we have not tried any correction to the original values. 
For the nine stars in common, [Fe/H]
from Edvardsson et al. are on average smaller by $0.07\pm 0.02$~dex (r.m.s. of
0.07 dex). Physical data for SWP are listed in Table~\ref{t:tab2} (only
available in electronic form).

\begin{table*}
\centering
\caption{Physical data for stars with planets (in electronic form)
References for metallicities
1 = \cite{san01}; 2 = CORAVEL {\tt http://obsunige.ch/}\~{\tt udry/planet/planet.html};
3 = \cite{gla00}; 4 = \cite{gon01}; 5 = \cite{gva98}; 6 = \cite{nae01};
7 = \cite{gon98}; 8 = \cite{gon99}; 9 = \cite{cay01}; 10 = \cite{v58};
11 = \cite{lgo01}
}
\scriptsize
\begin{tabular}{llllllllllll}
\hline
star             &    SpT &M$_{\rm v}$&[Fe/H]    &err.  &  ref.   & age   &  err.  & mass     & err.  \\
                 &        &          &          &      &         & Gyr   &        & M$_\odot$&       \\
\hline
HD 1237    &   G6   V     &   5.36   &  0.11    & 0.08 &  1      &  0.6  &        & 0.90   &         \\
HD 6434    &   G3   IV    &   4.69   & -0.55    & 0.07 &  1      & 12.4  &  1.0   & 0.86   &  0.01   \\
HD 8574    &   F8         &   3.90   & -0.09    &      &  2      &       &        & 1.10   &         \\
HD 9826    &   F8   V     &   3.47   &  0.12    & 0.05 &  3      &  2.8  &  0.2   & 1.28   &  0.04   \\
HD 10697   &   G5   IV    &   3.73   &  0.16    & 0.03 &  4      &  6.0  &        & 1.10   &  0.01   \\
HD 12661   &   G6   V     &   4.58   &  0.35    & 0.02 &  4      &  8.4  &        & 1.07   &         \\
HD 13445   &   K1   V     &   5.98   & -0.20    & 0.06 &  1      &  2.2  &        & 0.80   &         \\
HD 16141   &   G5   IV    &   4.00   &  0.15    & 0.05 &  1      &  8.5  &  0.5   & 1.01   &         \\
HD 17051   &   G0   V     &   4.22   &  0.25    & 0.06 &  1      &  1.6  &        & 1.03   &  0.02   \\
HD 19994   &   F8   V     &   3.32   &  0.26    & 0.06 &  1      &  3.1  &        & 1.32   &  0.05   \\
HD 22049   &   K2   V     &   6.20   & -0.07    & 0.06 &  1      &  1.0  &        & 0.75   &         \\
HD 27442   &   K2   IV    &   3.12   &  0.20    &      &  2      &       &        & 1.20   &         \\
HD 28185   &   G5         &   4.81   &  0.24    & 0.05 &  1      &       &        & 0.90   &         \\
HD 37124   &   G4   V     &   5.07   & -0.41    & 0.03 &  4      &  3.9  &        & 0.91   &         \\
HD 38529   &   G4   IV    &   2.80   &  0.39    & 0.06 &  1      &  3.0  &  0.5   & 1.39   &         \\
HD 46375   &   K1   IV-V  &   5.32   &  0.21    & 0.04 &  4      &  4.5  &        & 1.00   &   0.1   \\
HD 50554   &   F8         &   4.38   &  0.02    &      &  2      &       &        & 1.10   &         \\
HD 52265   &   G0   V     &   4.06   &  0.24    & 0.06 &  1      &  4.0  &        & 1.13   &         \\
HD 74156   &   G0         &   3.56   &  0.13    &      &  2      &       &        & 1.05   &         \\
HD 75289   &   G0   V     &   4.04   &  0.27    & 0.06 &  1      &  2.1  &  0.7   & 1.15   &  0.02   \\
HD 75732   &   G8   V     &   5.46   &  0.45    & 0.04 &  5      &  3.6  &  3.0   & 0.85   &         \\
HD 80606   &   G5         &   5.23   &  0.43    &      &  6      &       &        & 0.90   &         \\
HD 82943   &   G0         &   4.35   &  0.33    & 0.06 &  1      &  5.0  &        & 1.05   &         \\
HD 83443   &   K0   V     &   5.05   &  0.39    & 0.09 &  1      &       &        & 0.79   &         \\
HD 89744   &   F7   V     &   2.79   &  0.30    & 0.03 &  4      &  1.8  &  0.1   & 1.40   &  0.09   \\
HD 92788   &   G5   V     &   4.55   &  0.31    & 0.03 &  4      &  6.4  &        & 1.06   &         \\
HD 95128   &   G1   V     &   4.36   &  0.01    & 0.06 &  7      &  6.5  &  1.5   & 1.06   &  0.03   \\
HD 106252  &   G0         &   4.54   & -0.16    &      &  2      &  1.1  &        & 1.05   &         \\
HD 108147  &F8-G0   V     &   4.06   &  0.20    & 0.06 &  1      &  2.0  &        & 1.05   &         \\
HD 114762  &   F9   V     &   4.26   & -0.60    & 0.06 &  7      & 16.0  &        & 0.82   &  0.03   \\
HD 117176  &   G4   V     &   3.71   & -0.03    & 0.06 &  7      &  9.0  &        & 0.92   &         \\
HD 120136  &   F7   V     &   3.53   &  0.32    & 0.06 &  3      &  1.5  &  0.5   & 1.20   &         \\
HD 121504  &   G2   V     &   4.30   &  0.17    & 0.06 &  1      &  2.8  &        & 1.00   &         \\
HD 130322  &   K0   III   &   5.67   &  0.05    & 0.03 &  4      &  0.3  &        & 0.79   &         \\
HD 134987  &   G5   V     &   4.40   &  0.32    & 0.04 &  4      &  5.8  &        & 1.05   &         \\
HD 141937  &G2-G3   V     &   4.63   &  0.16    &      &  2      &  1.5  &        & 1.00   &         \\
HD 143761  &   G0   V     &   4.19   & -0.29    & 0.06 &  7      & 12.1  &  1.2   & 0.93   &  0.03   \\
HD 145675  &   K0   V     &   5.38   &  0.50    & 0.05 &  8      &  6.0  &        & 0.79   &         \\
HD 160691  &   G3   IV-V  &   4.23   &  0.28    &      &  9      &       &        & 1.08   &         \\
HD 168443  &   G6   IV    &   4.03   &  0.10    & 0.03 &  4      &  7.4  &        & 1.01   &  0.02   \\
HD 168746  &   G5         &   4.78   & -0.06    & 0.05 &  1      &       &        & 0.92   &         \\
HD 169830  &   F8   V     &   3.11   &  0.22    & 0.05 &  1      &  2.0  &  0.3   & 1.35   &  0.04   \\
HD 177830  &   K0   IV    &   3.32   &  0.36    & 0.05 &  4      & 13.5  &        & 1.15   &   0.2   \\
HD 178911  &   G5         &   4.62   &  0.06    & 0.02 &  9      &       &        & 0.90   &         \\
HD 179949  &   F8   V     &   4.10   &  0.00    &      & 10      &  3.3  &        & 1.24   &         \\
HD 186427  &   G3   V     &   4.60   &  0.07    & 0.03 & 11      &  7.0  &        & 1.00   &         \\
HD 187123  &   G3   V     &   4.43   &  0.16    & 0.05 &  8      &  4.0  &  1.0   & 1.00   &   0.1   \\
HD 190228  &   G5   IV    &   3.34   & -0.24    & 0.06 &  1      &       &        & 1.30   &         \\
HD 192263  &   K2   V     &   6.30   & -0.03    & 0.04 &  4      &  0.3  &        & 0.75   &         \\
HD 195019  &   G3   IV-V  &   4.01   & -0.12    &      & 10      &  9.5  &        & 0.98   &  0.06   \\
HD 209458  &   G0   V     &   4.28   &  0.04    & 0.03 &  4      &  4.3  &        & 1.10   &   0.1   \\
HD 210277  &   G7   V     &   4.99   &  0.23    & 0.05 &  1      &  6.9  &        & 0.92   &         \\
HD 213240  &   G4   IV    &   3.76   &  0.23    &      &  2      &  2.7  &        & 0.95   &         \\
HD 217014  &   G2   IV    &   4.56   &  0.21    & 0.03 &  4      &  7.0  &  1.0   & 1.10   &  0.04   \\
HD 217107  &   G8   IV    &   4.70   &  0.39    & 0.05 &  1      &  5.6  &        & 0.96   &         \\
HD 222582  &   G5   V     &   4.59   &  0.02    & 0.03 &  4      &  5.6  &        & 1.00   &         \\
BD-10 3166 &   K0   V     &          &  0.33    & 0.05 &  4      &  4.0  &        & 1.10   &   0.1   \\
GJ 876     &   M4   V     &   9.52   &          &      &         &       &        & 0.32   &  0.05   \\
\hline
\normalsize
\label{t:tab2}
\end{tabular}
\end{table*}

\section{Calculation of orbits}

We transformed the proper motions into the corresponding galactocentric velocity
components $\Pi, \Theta$, and $Z$, and corrected them for the Standard Solar
Motion and the Motion of the Local Standard of Rest (LSR). For the adjustment
of Standard Solar Motion we used a solar motion of $(U,V,W) = (+10.4,
+14.8, +7.3 )$ Km/s, according to \cite{mro68}. The adopted procedure follows
the method of \cite{jso87}; however we adopted a right-handed reference frame
with the $x$-axis pointing toward the anticenter. The $y$-axis is along the
direction of galactic rotation, and the $z$-axis is toward the North Galactic
Pole.

\subsection{Galactic model of mass distribution}

The equations of motion have been integrated adopting the model for the Galactic
gravitational potential and corresponding mass distribution by \cite{asa91}.

In this model, the mass distribution of the Galaxy is described as a three
component system: a spherical central bulge, and a flattened disk, both of the
Miyamoto-Nagai form, plus a massive spherical halo. The gravitational
potential is fully analytical, continuous everywhere, and has continuous
derivatives; its simple mathematical form lead to a rapid integration of the
orbits with high numerical precision. The model provides accurate
representation of the Galactic rotation curve $V_C(R)$ and the force $F_z(z)$
perpendicular to the Galactic Plane. The values obtained for the Galactic
rotation constants are $A= 12.95 ~\rm{Km}~\rm{s}^{-1}~\rm{kpc}^{-1}$ and
$B=-12.93~\rm{Km}~\rm{s}^{-1}~\rm{kpc}^{-1}$ which are in good agreement with
observational data.

The expression for the potential of three components is:
\begin{equation}
\phi_B(r,z) = - \frac{G M_B}{\sqrt{r^2+z^2+b_B^2}} 
\end{equation}

\begin{equation}
\phi_D(r,z) = - \frac{G M_D}{\sqrt{r^2+\bigg(a_D+\sqrt{z^2+b_D^2}\bigg)^2}} 
\end{equation}

\begin{displaymath}
\phi_H(r,z) = - \frac{G M_H}{\varrho}\cdot 
\frac{\big(\frac{\varrho}{a_H}\big)^{2.02}}{1+\big(\frac{\varrho}{a_H}\big)^{1.0
2}}
-\frac{M_H}{1.02\cdot a_H} \times
\end{displaymath}
\nonumber
\begin{equation}
\bigg[ -\frac{1.02}{1+\big(\frac{\varrho}{a_H}\big)^{1.02}} + 
\ln\Big(1+\Big(\frac{\varrho}{a_H}\Big)^{1.02} \Big) \bigg]^{100}_R 
\end{equation}
where $\varrho = \sqrt{r^2+z^2}$.

\noindent
Table \ref{t:tab3} lists the values of the various constants for this model.
The total mass of the model is $9.0\cdot 10^{11} {\rm M_\odot}$,
and the Halo is truncated at 100 kpc.

\begin{table}
\caption{Constants for the galactic model}
\begin{tabular}{lcrr}
\hline
        &       &                    &\\
Galactocentric distance of Sun & $R_\odot$ & 8.5   & kpc \\
Local circular velocity        & $\Theta$  & 220   & Km ${\rm s}^{-1}$ \\
        &       &                    &\\
Bulge   & $M_B$& 1.41$\cdot 10^{10}$ & ${\rm M}_\odot$ \\
        & $b_B$& 0.3873              & kpc \\
%        &       &                    &\\
Disk    & $M_D$& 8.56$\cdot 10^{10}$ & ${\rm M}_\odot$ \\
        & $a_D$& 5.3178              & kpc \\
        & $b_D$& 0.2500              & kpc \\
%        &       &                    &\\
Halo    & $M_H$& 80.02$\cdot 10^{10}$& ${\rm M}_\odot$ \\
        & $a_H$& 12.0                & kpc \\
        &       &                    &\\
\hline
\end{tabular}
\label{t:tab3}
\end{table}

\section{Galactic orbits}

To perform the numerical integration,
we utilized the Burlish-Stoer method, directly applied to the second order
differential equations that describe the motion of a star. This numerical
method allows to obtain a typical error in energy and in the $z$-component of
the angular momentum of the star of, respectively, $\Delta E/E \approx
10^{-4}$ and $\Delta L_z/L_z \approx 10^{-9}$.
Orbits were back integrated long enough to obtain significant values for the
main orbital parameters.

\subsection{Stars with planets}

We have not computed orbits for two stars:
BD-10$^\circ$~ 3166 because the parallax
is not known; and HD~222582 because the value of the heliocentric radial
velocity of this star is not available from the literature.

Galactocentric orbit parameters for the remaining 56 stars are given in
Table~\ref{t:tab4} (only available in electronic form), where $R_p$ is the
perigalacticon, $R_a$ is the apogalacticon, $z_{\rm max}$ is the maximum height 
above the Galactic Plane, $e$ is the eccentricity defined as $\frac{R_a - R_p}{R_a
+R_p}$, $E$ is the total energy, $L_z$ is the $z$-component of the angular
momentum of the stars.

\begin{table*}
\centering
\caption{Galactic Orbit Parameters for Stars with Planets (in electronic form)}
\scriptsize
\begin{tabular}{lcccccccccc}
\hline
   Star       &    $R_p$     &    $R_a$    &   $z_{max}$ &      $e$   &     $E$     &  $L_z$    \\
              &    kpc       &    kpc      &   kpc       &            &${\rm km}^2\cdot {\rm s}^{-2}$&${\rm kpc}\cdot {\rm km}\cdot {\rm s}^{-1}$\\
\hline
  HD  1237    &        8.01  &       8.95  &       0.12  &       0.06 &  -111864.3  &    1878.8 \\
  HD  6434    &        5.09  &       9.92  &       0.04  &       0.32 &  -118449.9  &    1511.1 \\
  HD  8574    &        7.04  &       8.94  &       0.25  &       0.12 &  -114919.4  &    1749.7 \\
  HD  9826    &        7.27  &       9.30  &       0.04  &       0.12 &  -113183.3  &    1816.7 \\
  HD  10697   &        6.93  &       9.44  &       0.34  &       0.15 &  -113457.0  &    1775.3 \\
  HD  12661   &        6.44  &       9.87  &       0.15  &       0.21 &  -113919.2  &    1740.7 \\
  HD  13445   &        4.82  &       9.50  &       0.33  &       0.33 &  -120730.8  &    1428.5 \\
  HD  16141   &        5.87  &      10.62  &       0.18  &       0.29 &  -113232.9  &    1696.7 \\
  HD  17051   &        7.99  &       8.87  &       0.02  &       0.05 &  -112239.7  &    1869.4 \\
  HD  19994   &        8.09  &       8.65  &       0.02  &       0.03 &  -112628.4  &    1858.8 \\
  HD  22049   &        8.49  &      10.06  &       0.15  &       0.09 &  -107204.0  &    2044.3 \\
  HD  27442   &        7.50  &       8.89  &       0.14  &       0.09 &  -113715.1  &    1808.1 \\
  HD  28185   &        7.00  &       8.80  &       0.20  &       0.11 &  -115640.2  &    1732.8 \\
  HD  37124   &        6.47  &       9.02  &       0.39  &       0.16 &  -116354.9  &    1670.4 \\
  HD  38529   &        7.77  &       8.56  &       0.27  &       0.05 &  -113687.5  &    1806.2 \\
  HD  46375   &        7.87  &       8.79  &       0.17  &       0.06 &  -112750.6  &    1845.2 \\
  HD  50554   &        8.28  &       9.14  &       0.03  &       0.05 &  -110501.3  &    1931.4 \\
  HD  52265   &        7.17  &       9.38  &       0.01  &       0.13 &  -113217.9  &    1811.1 \\
  HD  74156   &        6.20  &       8.91  &       0.10  &       0.18 &  -118253.6  &    1628.8 \\
  HD  75289   &        7.79  &       9.33  &       0.13  &       0.09 &  -111341.4  &    1888.4 \\
  HD  75732   &        7.73  &       9.07  &       0.04  &       0.08 &  -112428.9  &    1856.6 \\
  HD  80606   &        8.41  &      10.00  &       0.23  &       0.09 &  -107486.4  &    2027.3 \\
  HD  82943   &        7.79  &       8.88  &       0.01  &       0.07 &  -112839.6  &    1846.4 \\
  HD  83443   &        6.96  &       8.87  &       0.02  &       0.12 &  -115716.3  &    1736.0 \\
  HD  89744   &        7.60  &       8.54  &       0.07  &       0.06 &  -114602.6  &    1788.4 \\
  HD  92788   &        7.63  &       9.01  &       0.12  &       0.08 &  -112889.5  &    1837.3 \\
  HD  95128   &        8.41  &       9.47  &       0.12  &       0.06 &  -109072.5  &    1979.2 \\
  HD  106252  &        6.51  &       8.93  &       0.12  &       0.16 &  -117018.1  &    1676.7 \\
  HD  108147  &        8.20  &       9.06  &       0.05  &       0.05 &  -110951.1  &    1914.6 \\
  HD  114762  &        5.43  &       9.28  &       1.25  &       0.26 &  -116978.9  &    1488.1 \\
  HD  117176  &        6.30  &       8.65  &       0.06  &       0.16 &  -118934.1  &    1622.4 \\
  HD  120136  &        7.88  &       8.90  &       0.02  &       0.06 &  -112506.6  &    1858.5 \\
  HD  121504  &        6.23  &       8.50  &       0.07  &       0.15 &  -119794.1  &    1598.0 \\
  HD  130322  &        7.79  &       8.48  &       0.04  &       0.04 &  -114188.3  &    1805.2 \\
  HD  134987  &        7.01  &       8.49  &       0.32  &       0.10 &  -116481.3  &    1701.3 \\
  HD  141937  &        8.41  &      10.58  &       0.02  &       0.11 &  -106064.5  &    2084.1 \\
  HD  143761  &        6.60  &       9.76  &       0.37  &       0.19 &  -113409.1  &    1752.0 \\
  HD  145675  &        7.77  &       9.41  &       0.10  &       0.09 &  -111145.3  &    1895.6 \\
  HD  160691  &        8.48  &       8.93  &       0.05  &       0.03 &  -110533.7  &    1932.8 \\
  HD  168443  &        5.83  &       8.58  &       0.02  &       0.19 &  -121050.8  &    1544.8 \\
  HD  168746  &        7.78  &       8.55  &       0.05  &       0.05 &  -113970.3  &    1811.4 \\
  HD  169830  &        8.42  &       9.23  &       0.10  &       0.05 &  -109775.5  &    1956.3 \\
  HD  177830  &        5.00  &       8.52  &       0.01  &       0.26 &  -124689.1  &    1402.4 \\
  HD  178911  &        7.08  &       9.41  &       0.08  &       0.14 &  -113386.1  &    1800.4 \\
  HD  179949  &        8.06  &       8.96  &       0.01  &       0.05 &  -111733.5  &    1887.0 \\
  HD  186427  &        7.08  &       8.80  &       0.05  &       0.11 &  -115542.8  &    1745.3 \\
  HD  187123  &        8.11  &       8.70  &       0.38  &       0.03 &  -111935.1  &    1856.3 \\
  HD  190228  &        6.20  &       8.51  &       0.27  &       0.16 &  -119545.0  &    1590.5 \\
  HD  192263  &        8.47  &      10.25  &       0.34  &       0.09 &  -106502.2  &    2055.2 \\
  HD  195019  &        4.44  &       9.05  &       0.26  &       0.34 &  -124305.1  &    1328.1 \\
  HD  209458  &        8.25  &       8.54  &       0.10  &       0.02 &  -112419.8  &    1865.0 \\
  HD  210277  &        6.34  &       8.54  &       0.05  &       0.15 &  -119196.5  &    1619.2 \\
  HD  213240  &        7.18  &       9.06  &       0.35  &       0.12 &  -113875.5  &    1775.8 \\
  HD  217014  &        7.32  &       8.50  &       0.29  &       0.07 &  -115385.8  &    1744.0 \\
  HD  217107  &        8.37  &       8.95  &       0.22  &       0.03 &  -110614.9  &    1919.8 \\
  GJ  876     &        8.16  &       8.50  &       0.02  &       0.02 &  -112896.6  &    1850.8 \\
\hline
\end{tabular}
\normalsize
\label{t:tab4}
\end{table*}

\subsection{Stars of Edvardsson catalog}

The Edvardsson catalog contains 189 stars, but we have computed the orbits for
185 stars; radial velocities are not available in the literature for the four 
remaining stars (HD 98553, HD 155358, HD 159703, HD 218504).

We have recalculated the galactic orbit of stars of the Edvardsson catalog,
although the data about perigalacticon are present in this catalog, because
Edvardsson assumed a solar galactocentric distance of 8.0 kpc rather than the
value of 8.5 kpc adopted in this paper \citep{klb86}.

\section{Results}

To understand whether the high metallicity of SWP is the cause or the effect of
the presence of planets, we plotted the iron abundance relative to
perigalacticon for both samples. 

\begin{figure}
\includegraphics[angle=-90, width=8.8cm]{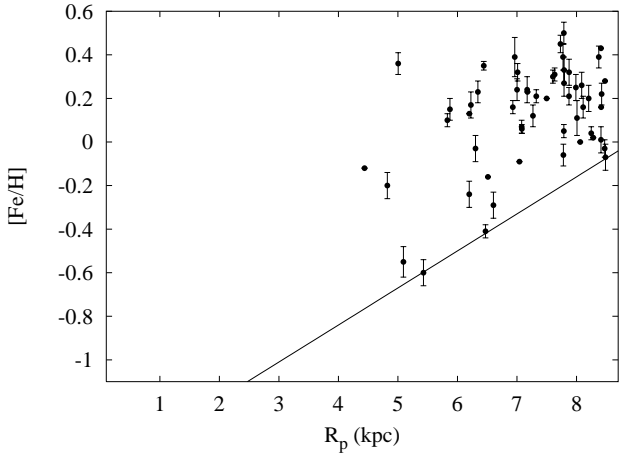}
\caption[ ]{ Perigalactic distances for the SWP sample,
the straight line represent the lower envelope of SWP distribution in this plane}
\label{f:fig4}
\end{figure}

\begin{figure}
\includegraphics[angle=-90, width=8.8cm]{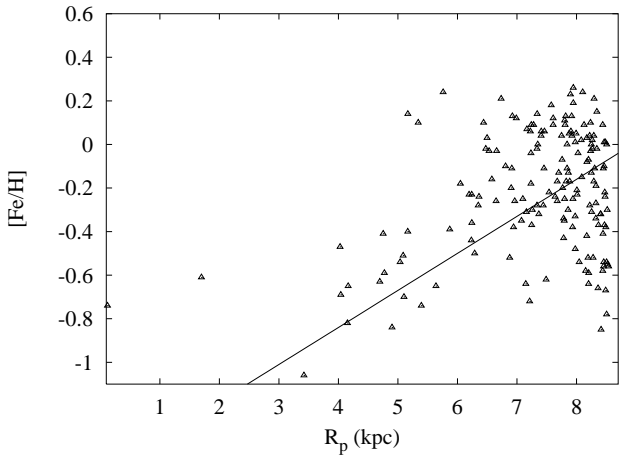}
\caption[ ]{ Perigalactic distances for the Edvardsson sample;
the straight line is the lower envelope of the SWP distribution}
\label{f:fig5}
\end{figure}

Figures~\ref{f:fig4} and \ref{f:fig5} give the [Fe/H] versus
perigalacticon for the two samples. These figures show that the distribution
of metallicity versus perigalacticon of the two samples are quite similar: in
both cases metallicity increases with perigalactic distance. To understand this
trend (apparently opposite to the overall radial abundance gradient found
for our Galaxy using various techniques), we note that our samples
are local: stars with small perigalactic distances that presently are close to
the Sun should be on highly eccentric orbits; they are drawn from the thick
disk or old thin disk populations, and are on average much older (and
metal-poor) than the stars on more circular orbits. Practically, our data
indicates the presence of two different populations of SWP: an old population
with perigalactic distance less than 6 kpc with lower metallicity, and a young
population with perigalactic distance more than 6 kpc and with high
metallicity.

\section{Discussion}

Our data may be used to discuss the metallicity-planet connection. In fact, if
the high metallicity is the cause for the presence of planets, we should not
expect any correlation between presence of planets and galactocentric
distance: it is the overall metallicity that is important, and in a given
metallicity bin the distribution of stars with perigalactic distance should be
the same for stars with and without planets. On the other hand, if planet
capture is the mechanism that enhances the metallicity, we should expect that
the SWP are the upper envelope of the distribution of metallicity with
galactocentric distance: at any galactocentric distance the stars with planets
should be more metal-rich than average, and in a given metallicity bin, stars
with planets should have on average smaller perigalactic distances.

At any galactocentric distance, the metallicity of the stars with planets
is roughly the upper envelope of the metallicity distribution of the comparison
sample. The few known (mildly) metal-poor stars with planets all have small 
values of
the perigalactic distance, and there are no stars with known planets among
the much more frequent (in the solar neighborhood) mildly metal-poor stars
that have nearly circular orbits. Based on the relative frequency in the
solar neighborhood of mildly metal-poor stars on circular and highly eccentric
orbits, we should expect a large number of mildly metal-poor stars with planets.
These are not observed.

A Kolmogorov-Smirnov test for the mildly metal-poor ([Fe/H] $\le -0.1$)
SWP and Edvardsson stars gives a probability that both distributions with
perigalactic distances were extracted from the same parent population of
0.0046; the same test for the higher value of metallicity ([Fe/H] $>
-0.1$) gives a probability of 0.5159. This result does not depend
critically on possible offsets between the abundance scales used for SWP and
the Edvardsson sample: in fact, even if we lower [Fe/H] for all SWP by
0.07~dex (the systematic offset measured for the nine stars in common in the
two samples), the possibility that the mildly metal-poor stars ([Fe/H] $\le
-0.1$) in the two samples were extracted from parent populations having the
same distribution with perigalactic distance can still be rejected at 
a quite good level of confidence (the probability that they are drawn from 
the same distribution is 0.022). The test for
the higher metallicity objects gives a probability of 0.6644.

Note that this is not due to a selection effect related to the
apparent brightness of the stars: in fact in the Edvardsson sample, metal-poor
stars ([Fe/H] $\le -0.1$) with $R_p \le 7$ kpc are on average fainter than metal-poor
stars with $R_p > 7$ kpc (average magnitudes of 6.16 $\pm$ 0.20 and 5.80 $\pm$ 0.12
respectively).
It should then be easier to discover planets around the brighter stars 
with near circular orbits than around the fainter stars with eccentric orbits.

In the scenario where high metallicity is the cause of the presence of 
planets, when the metallicity exceeds a critical value, the planets are 
present, independent of the perigalactic distance. 
Hence we expect that in a given metal abundance range, the number of stars with 
planets at different perigalactic distance will follow the distribution of the 
parent population of all nearby stars. We should then find more stars with 
planets with large perigalactic distance than with small perigalactic distance 
(since the parent population is more numerous).

In the opposite scenario, where the planets are the cause of the high 
metallicity of the central stars, the SWP are the upper envelope 
of the distribution of [Fe/H] versus perigalactic distance of field stars: 
this is because at any perigalactic distance, stars with planets are metal 
enriched with respect to stars without planets. In this case we expect that 
in the mildly metal-poor abundance bin, we may find a larger number of stars 
with small perigalactic distance than stars with large perigalactic distance, 
because the metal enrichment caused by the presence of the planets will push
the stars with planets extracted from the parent population of mildly 
metal-poor stars out of this metallicity bin (into the high metallicity bin). 

We conclude that insofar as no kinematic selection effect is present in 
the sample of stars with planets, the fact that planets have been found 
only among mildly metal-poor stars (presently in the solar
neighborhood) that have small perigalactic radii (in spite of the 
relative rarity of such objects) clearly favours the second scenario.

We argue that this result strongly favours scenarios where the presence of
planets is the cause of the higher metallicities.
Of course our results do not rule out the possibility that higher metallicity
also favours the presence of planets (i.e. that both scenarios are applicable).

\acknowledgements{ This research has made use of the SIMBAD data base,
operated at CDS, Strasbourg, France. \\
We thanks Dr. S.J. Aarseth for making available his code ORBIT, 
Dr. G. Carraro for useful discussion and suggestions,
Dr. S. Desidera for a critical reading of the text,
Dr. G. Marcy and Dr M. Mayor for prompt responses to our inquiry about
any possible kinematic bias in their sample, and
the referee Dr. M. Cr\'ez\'e that helped to improv our paper.
}

\end{document}